# Strong thermoelectric response of nanoconfined weak electrolytes

Rajkumar Sarma and Steffen Hardt*

Technische Universität Darmstadt, Fachbereich Maschinenbau, Fachgebiet Nano- und Mikrofluidik

*hardt@nmf.tu-darmstadt.de

**Abstract:** When dissolved, weak electrolytes only partially dissociate into ions in a temperature-dependent process. We show herein that such incomplete dissociation yields an enormous thermoelectric response in an electrolyte-filled nanochannel along which a temperature gradient is applied. For this purpose, an extended version of the Nernst-Planck equations is developed that takes into account the temperature-dependent dissociation-association equilibrium. The results indicate that in this way, Seebeck coefficients can be achieved that outperform all previously reported values.

In recent years, the recovery of low-grade waste heat has moved into the focus of intense research activities [1,2]. Large amounts of waste heat from the industrial, transportation or residential sector are usually lost to the environment. Owing to their cost and effort of implementation, solid-state thermoelectric devices, which partially convert a heat flux into electric power, often do not represent a viable option to recover a part of this energy. In the past few years, ionic thermoelectric materials have gained increasing attention [3–7]. These are often composed of a nanoporous solid matrix filled with a liquid electrolyte. They not only outperform their solid-state counterparts in terms of the thermovoltage achieved, but also in terms of cost, non-toxicity and scalability.

The performance of thermoelectric materials is often characterized using the Seebeck coefficient $\mathcal{S}$, which measures the thermovoltage per applied temperature gradient, according to $\Delta V = \mathcal{S}\Delta T$. Very high Seebeck coefficients of ionic thermoelectric materials of the order of 10 mV/K or higher have been achieved with ionic thermoelectric materials (see, e.g., [8–11]). These materials are often very complex, which makes it difficult to pinpoint the physics behind the large thermoelectric response. One overarching feature of many materials is their nanoscale structures. The physics of thermoelectricity in nanochannels filled with electrolyte solutions was studied in [12]. These findings show how nanoconfinement augments thermoelectricity, but are probably not sufficient to explain the very large Seebeck coefficients found in some experiments. Experimental studies suggest that complex electrolytes such as polyelectrolytes or ionic liquids seem to promote large thermoelectric responses. We have recently shown [13] that the specific physics of charge carrier generation in nanoconfined ionic liquids, which is related to an unexpectedly large Debye length [14,15], can yield augmented Seebeck coefficients. In these cases, charge carrier concentration gradients in the overlapping electric double layers (EDLs) of a nanochannel drive thermoelectricity. While the corresponding physics can usually be understood using continuum mechanical models, thermoelectricity can also be induced by effects that require a description on the molecular level. For example, thermoelectric currents can be driven by gradients in the excess enthalpy of a liquid very close to a channel wall [16] or by thermally activated hopping of charge carriers between the minima of a potential energy landscape [17].

Here, we report that in a confined domain, weak electrolytes yield Seebeck coefficients that can be larger than anything previously reported in the literature. Unlike strong electrolytes, weak electrolytes dissociate into their constituent ions only to a small degree, which is characterized by the temperature-dependent equilibrium dissociation constant $\mathcal{K}_{eq}$ [18–22]. To study the thermoelectric response of nanoconfined weak electrolytes, we present extended Poisson-Nernst-Planck (PNP) equations that take into account the chemical equilibrium between dissociated and un-dissociated solutes. These equations are solved analytically and numerically to compute the thermovoltage across an electrolyte-filled nanochannel with an axial temperature gradient.

Figure 1a schematically depicts the system we consider. A slit nanochannel filled with a weak electrolyte solution connects two reservoirs maintained at temperatures $T_0$ and $T_e$, respectively. The temperature difference $\Delta T\left(=T_e-T_0>0\right)$ between the reservoirs establishes a temperature gradient $\nabla T$ within the system. The charged channel walls exhibit EDLs, which overlap if the channel width is small enough. In that case, the channel is mainly filled with counter-ions. Due to the temperature-dependent equilibrium dissociation constant $\mathcal{K}_{eq}=\mathcal{K}_{eq}\left(T\right)$, the axial temperature gradient translates into corresponding concentration gradients of neutral solute, co- and counter-ions (cf. Fig. 1b). This relates to the fact that the Debye length $\kappa^{-1}$ varies along the channel owing to the temperature-dependent charge carrier concentration. The axial concentration gradient drives ion transport through the channel. To simplify the analysis, we consider charge transport only through the diffuse layer and ignore the potential impact of a Stern layer.

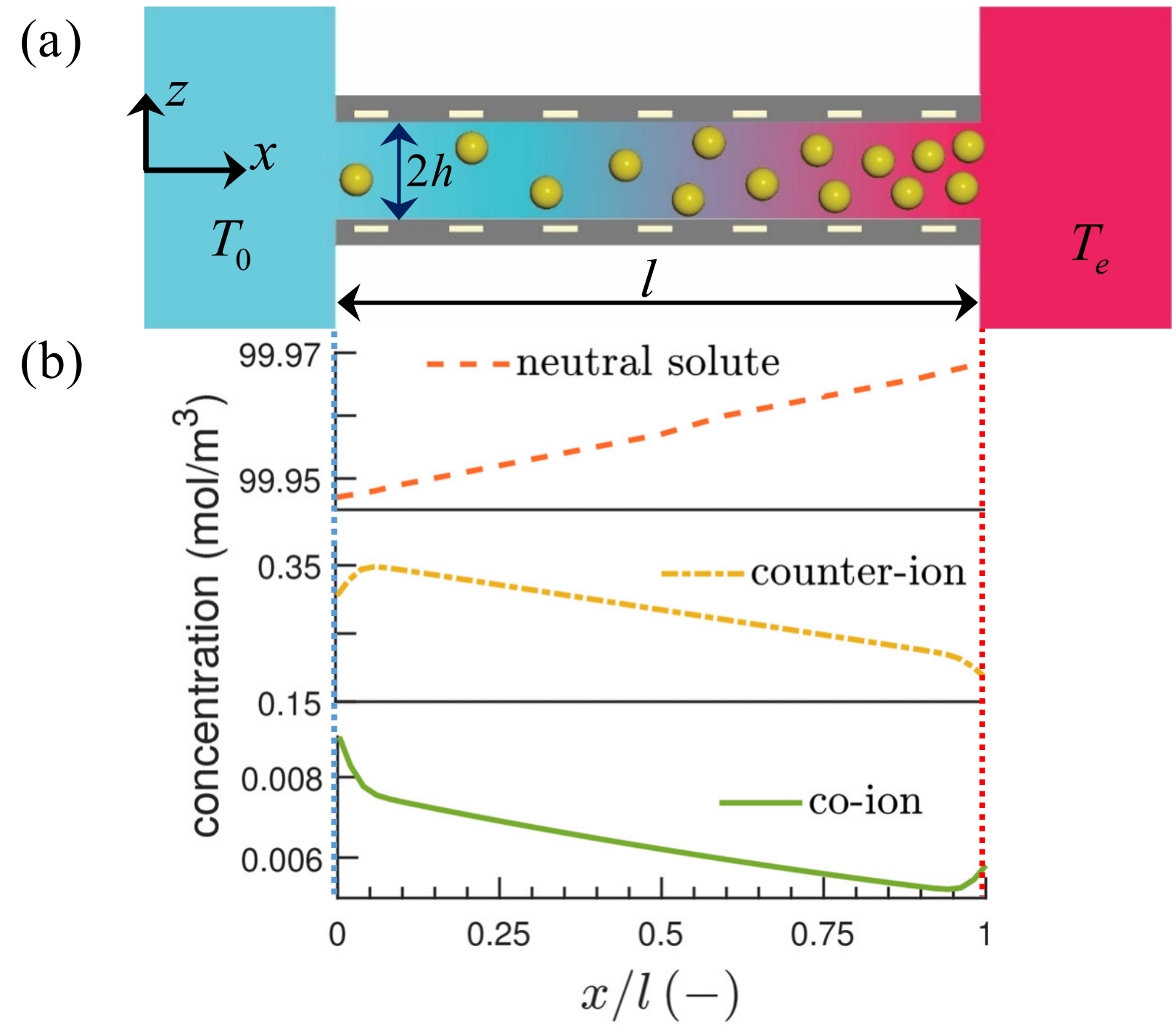

**FIG. 1.** Schematic illustration of the physical system. (a) A weak electrolyte solution is confined between two parallel plates with a gap width of $2h$, connecting two reservoirs maintained at temperatures $T_0$ and $T_e\left(=T_0+\Delta T\right)$, respectively. Each plate is kept at a constant $\zeta$ potential, leading to the development of electrical double layers of thickness $\kappa^{-1}$ in its close vicinity. The temperature-dependent $\mathcal{K}_{eq}$ establishes an axial concentration gradient of the constituent species within the confinement. (b) Cross-sectional area averaged concentration of each constituent species in the axial direction for $\kappa_0 h=0.1$, $\zeta=-50\,\mathrm{mV}$, $T_0=293\,\mathrm{K}$, and $\Delta T=30\,\mathrm{K}$.

A partial dissociation of the ionizing solutes suggests the presence of three distinct solutes in a weak electrolyte solution: cations, anions and neutral solutes. For weak electrolytes, the standard PNP model pertinent to the strong electrolytes needs to get modified to take into account the temperature-dependent dissociation-association reaction. We denote the rate constants for the dissociation and association reactions as $\mathcal{K}_d$ and $\mathcal{K}_a$, respectively. In the spirit of a proof-of-principle study, we consider a symmetric weak electrolyte solution with only one type of neutral solute, cation and anion with valence 0, $+\nu$ and $-\nu$, respectively. The species have identical diffusivities $D$, translating to equal electrophoretic mobilities $\mu\left(=D/k_BT\right)$ of the cations and anions, as given by the Einstein-Smoluchowski relation. It is shown in the supplemental material [23] that, for the transport processes we are analyzing, advective transport is negligible. Also, to highlight the effects due to nanoconfinement, we have neglected any species transport due to their intrinsic thermophoretic mobility. Therefore, the Nernst-Planck equations, describing the evolution of the concentration fields $c_k$, read

$$-\nabla\cdot\left(D\nabla c_k + e\nu_k\mu c_k\nabla\phi\right) = \mathcal{G}_k , \quad (1)$$

where $k=n$ represents the neutral solute, $k=+$ stands for the cation, and $k=-$ for the anion. In Eq. (1), the reaction term $\mathcal{G}_k$ describes the formation of the $k^{\mathrm{th}}$ species due to the dissociation-association reaction. In terms of the overall reaction rate $\mathcal{R}$, $\mathcal{G}_k$ can be expressed as $\mathcal{G}_k=\delta_k\mathcal{R}$, where $\mathcal{R}=\mathcal{K}_d c_n-\mathcal{K}_a c_+ c_-$. We have $\delta_n=-1$ and $\delta_+=\delta_-=1$. In the supplemental material [23] it is shown that the dissociation-association reaction is usually fast enough to assume local chemical equilibrium. Under this assumption, $\mathcal{K}_{eq}=c_+c_-/c_n$, and we denote $c_n+c_i=c_b$, $i=\{+,-\}$ with $c_b$ being independent of *i*. Furthermore, in Eq. (1), $e$ is the elementary charge and $\nu_k$ is the valence of species *k*. The total electric potential $\phi=\psi+\varphi$ is composed of the EDL potential $\psi$ and the induced thermoelectric potential $\varphi$, so that $E=-d\varphi/dx$ represents the induced thermoelectric field. $\phi$ satisfies the Poisson equation $\nabla\cdot\left(\epsilon\nabla\phi\right)=-\rho_f$, with $\rho_f=\sum_k e\nu_k c_k$ denoting the volumetric charge density and $\epsilon=\epsilon(T)$ the temperature-dependent dielectric permittivity.

An order-of-magnitude analysis reveals that the effects of advection and viscous dissipation can be neglected in the energy equation. Hence, as shown in [23], the temperature distribution becomes $T=T_0+\left(x/l\right)\Delta T$ inside the channel, which is further verified by the numerical simulations discussed below. One can compute $E$ by setting the net electric current $I=e\int_0^h\sum_k\nu_k\left(D\nabla c_k+e\nu_k\mu c_k\nabla\phi\right)dz$ to zero. This results in

$$E=\left[\int_0^h\left(\nabla c_+-\nabla c_-\right)dz+\frac{e\nu}{k_BT}\int_0^h\left(c_+\nabla\psi+c_-\nabla\psi\right)dz\right]\Bigg/\frac{e\nu}{k_BT}\int_0^h\left(c_++c_-\right)dz . \quad (2)$$

In the following, the Seebeck coefficient $\mathcal{S}=E/\nabla T$ is computed using both a numerical and an analytical approach.

For calculating $\mathcal{S}$ analytically, we first need to determine the local distributions of $\psi$ and $c_k$. Our approach is based on the leading-order contributions to the transport equations (except for the energy equation) in the small parameter $h/l\left(=A\right)\ll 1$, with *h* and *l* representing the half-width and length of the channel, respectively. Solving Eq. (1) in terms of the leading-order contribution in *A* [23], we obtain

$$c_i = c_0 \exp\left(-e\nu_i\psi/k_B T\right), \tag{3}$$

where $i=\left(+,-\right)$, and $c_0\left(=\sqrt{c_n\mathcal{K}_{eq}}\right)$ refers to the concentration of either charged species in the electroneutral $\left(\psi=0\right)$ region. To compute $\psi$, we proceed by substituting $c_i$ given by Eq. (3) into the Poisson equation. At leading order in *A*, the Poisson equation takes the form $d^2\bar{\psi}/dz^2=\kappa^2\sinh\left(\bar{\psi}\right)$, where $\bar{\psi}=e\nu\psi/k_BT$ and $\kappa^2=2e^2\nu^2\sqrt{\mathcal{K}_{eq}c_n}/\epsilon k_BT$ [23]. Note that, in the presence of an axial temperature gradient, the local Debye parameter $\kappa$ and accordingly $\psi$ are functions of *x*. Imposing the boundary conditions $\psi=\zeta$ at $z=\pm h$, the EDL potential under the Debye-Hückel approximation reads $\psi=\zeta\cosh\left(\kappa z\right)/\cosh\left(\kappa h\right)$. In this limit, the Seebeck coefficient is obtained as [23]

$$\mathcal{S}=\frac{-\dfrac{1}{\mathcal{K}_{eq}\left(2+\sqrt{\mathcal{K}_{eq}/c_n}\right)}\dfrac{d\mathcal{K}_{eq}}{dT}\zeta\dfrac{\tanh\left(\kappa h\right)}{\kappa h}}{1+\left(\dfrac{e\nu\zeta}{2k_BT}\right)^2\left[\dfrac{\tanh\left(\kappa h\right)}{\kappa h}+\dfrac{1}{\cosh^2\left(\kappa h\right)}\right]}+\frac{\dfrac{\zeta}{T}\dfrac{\tanh\left(\kappa h\right)}{\kappa h}\left\{1+\dfrac{1}{2}\left(\dfrac{e\nu\zeta}{k_BT}\right)^2\left[\dfrac{\tanh^2\left(\kappa h\right)}{3}+\dfrac{1}{\cosh^2\left(\kappa h\right)}\right]\right\}}{1+\left(\dfrac{e\nu\zeta}{2k_BT}\right)^2\left[\dfrac{\tanh\left(\kappa h\right)}{\kappa h}+\dfrac{1}{\cosh^2\left(\kappa h\right)}\right]}, \tag{4}$$

where $T=T_0+\Delta T$. For small $\zeta$ ($\left|e\nu\zeta/k_BT\right|<1$), the contribution of the terms $O\left(\zeta^2\right)$ can be ignored in Eq. (4). This results in

$$\mathcal{S}=\zeta\left[-\frac{1}{\mathcal{K}_{eq}\left(2+\sqrt{\mathcal{K}_{eq}/c_n}\right)}\frac{d\mathcal{K}_{eq}}{dT}+\frac{1}{T}\right]\frac{\tanh\left(\kappa h\right)}{\kappa h}. \tag{5}$$

From Eq. (5), it becomes clear that the thermoelectric field in a confined weak electrolyte solution is due to a superposition of two distinct mechanisms. The first one is an ion concentration gradient emanating from the temperature dependence of the dissociation constant, i.e., $d\mathcal{K}_{eq}/dT$. The second one is the concentration gradient originating from the temperature dependence of $D/\mu$, the parameter that determines the Debye length for a given ion concentration. For fully dissociated electrolytes ($\mathcal{K}_{eq} \to \infty$), only the latter term is responsible for the thermoelectric field [12]. In the forthcoming analysis it will be shown that the term proportional to $d\mathcal{K}_{eq}/dT$ can yield a much stronger thermoelectric response than the term proportional to $1/T$.

$\mathcal{S}$ attains a maximum when $\kappa h \to 0$ and vanishes for $\kappa h \to \infty$. This implies that the induced thermoelectric potential is at its maximum when the non-dimensional channel width is minimal (or the degree of EDL overlap is maximal). According to experiments with weak electrolytes, $\mathcal{K}_{eq} \sim O\left(10^{-3} - 10^{-16}\right)\mathrm{M}$ and $d\mathcal{K}_{eq}/dT \sim O\left(10^{-5} - 10^{-16}\right)\mathrm{M/K}$ [24]. Notably, qualitatively different forms of the function $\mathcal{K}_{eq}(T)$ have been reported. For example, for picric acid in 2-methoxyethanol, $\mathcal{K}_{eq}$ monotonically decreases with $T$, i.e., $d\mathcal{K}_{eq}/dT < 0$ [18]. On the other hand, acetic acid and formic acid in potassium hydroxide display a non-monotonic variation of the dissociation constant with temperature, leading to $d\mathcal{K}_{eq}/dT < 0$ or $> 0$, depending on the temperature range [25,26]. Accordingly, the ion flux induced by the temperature dependence of $\mathcal{K}_{eq}$ may either be along or against the temperature gradient. In total, the sign of the Seebeck coefficient is determined by the competition of the two terms in brackets of Eq. (5).

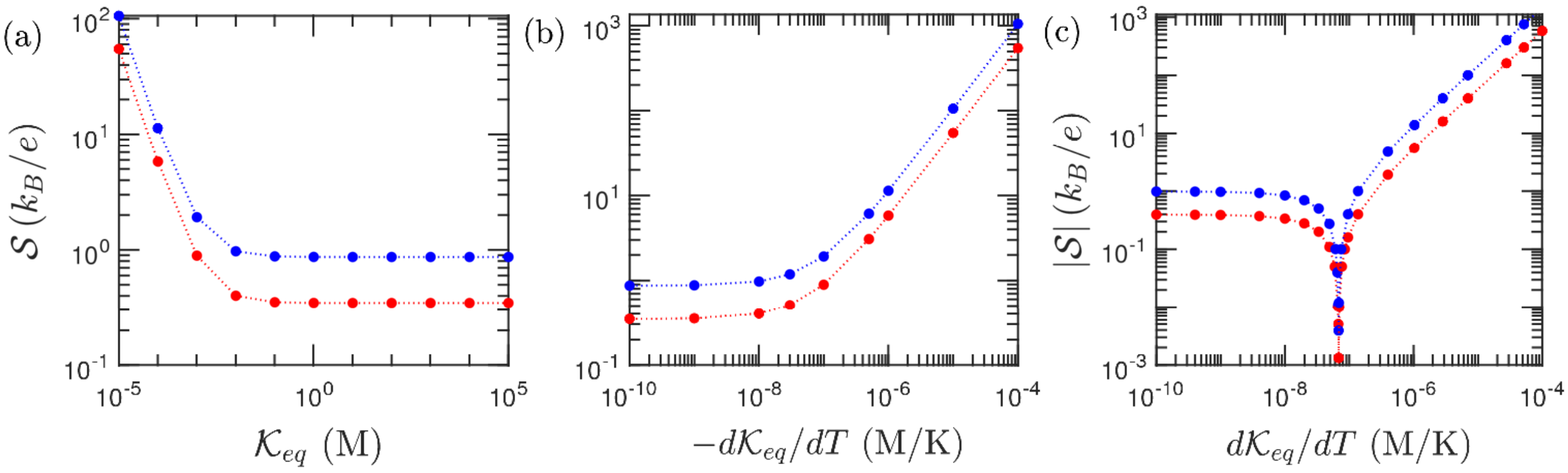

**FIG. 2.** Seebeck coefficient (in units of $k_B/e$) as a function of (a) the equilibrium dissociation constant $\mathcal{K}_{eq}$ and (b,c) its temperature sensitivity ($\pm d\mathcal{K}_{eq}/dT$). The symbols represent results from Eq. (5); the lines are guides to the eye. In each panel, the red symbols correspond to $\zeta = 10\,\mathrm{mV}$, while the blue symbols stand for $\zeta = 25\,\mathrm{mV}$. In panel (a), $d\mathcal{K}_{eq}/dT = -1\times10^{-5}$ M/K; in panel (b,c) $\mathcal{K}_{eq} = 1\times10^{-5}\,\mathrm{M}$. All results were obtained for $T = 293\,\mathrm{K}$ and $\kappa h = 0.1$.

$\mathcal{S}$ given by Eq. (5) is plotted as a function of $\mathcal{K}_{eq}$ in Fig. 2a and in terms of $d\mathcal{K}_{eq}/dT$ in Figs. 2(b,c) for a strongly confined electrolyte solution. One can observe that both $\mathcal{K}_{eq}$ and $d\mathcal{K}_{eq}/dT$ significantly impact the induced thermoelectric field, yielding Seebeck coefficients of the order of $10^3$ $k_B/e$. Translating this to the system of unit that is often used in the experimental literature, this results in values of 80 – 90 mV/K, which is larger than any value previously reported in the literature, to the best of our knowledge. Note that the range of $\mathcal{K}_{eq}$ and $d\mathcal{K}_{eq}/dT$ values was chosen in accordance with the literature [18,19,24,27,28]. The smaller the value of $\mathcal{K}_{eq}$, the stronger the thermoelectric field, as can be seen in Fig. 2(a). As $\mathcal{K}_{eq}$ increases, $\mathcal{S}$ decreases and reaches an asymptotic value characteristic for fully dissociated electrolytes. This indicates that, for weak electrolytes with sufficiently low values of $\mathcal{K}_{eq}$, the ion flux originating from the temperature-dependent partial dissociation of neutral solutes dominates over the ion flux resulting from the temperature dependence of $D/\mu$. Figs. 2(b,c) demonstrate that the higher the temperature sensitivity of $\mathcal{K}_{eq}$, the higher the magnitude of $\mathcal{S}$. Note that, for electrolytes with $d\mathcal{K}_{eq}/dT > 0$, only the magnitude of $\mathcal{S}$ is plotted in Fig. 2c. This is because of the sign reversal of $\mathcal{S}$ beyond a critical value of $d\mathcal{K}_{eq}/dT$. In order to understand the reason behind this sign change, one has to look into the underlying mechanisms contributing to the thermoelectric field. For $d\mathcal{K}_{eq}/dT < 0$, the ion fluxes due to the temperature dependence of $\mathcal{K}_{eq}$ and of $D/\mu$ are in the same direction, inducing a thermoelectric field whose magnitude monotonically increases with increasing temperature sensitivity of the weak electrolyte. However, the associated fluxes counteract when $d\mathcal{K}_{eq}/dT > 0$, leading to the disappearance of the thermoelectric field at a critical value of $d\mathcal{K}_{eq}/dT$ and causing a sign switch of $\mathcal{S}$ beyond it. For a small temperature sensitivity of $\mathcal{K}_{eq}$, again the Seebeck coefficient corresponding to a fully dissociated electrolyte is recovered. Note that for most weak electrolyte solutions, $\mathcal{K}_{eq}$ displays a nonlinear variation with temperature [24–26]. Therefore, while for strong electrolytes, the Seebeck coefficient usually does not depend on $\Delta T$, for weak electrolytes we generally expect a $\Delta T$-dependence. This is discussed in [23].

Up to this point, our analysis was about general weak electrolyte systems characterized by $\mathcal{K}_{eq}$ and $d\mathcal{K}_{eq}/dT$. Despite the vast variety of weak electrolytes, experimental data for dissociation constants as a function of temperature is scarce. To compute the Seebeck coefficient achievable with a specific system, we consider the weak electrolyte solution picric acid in 2-methoxyethanol.

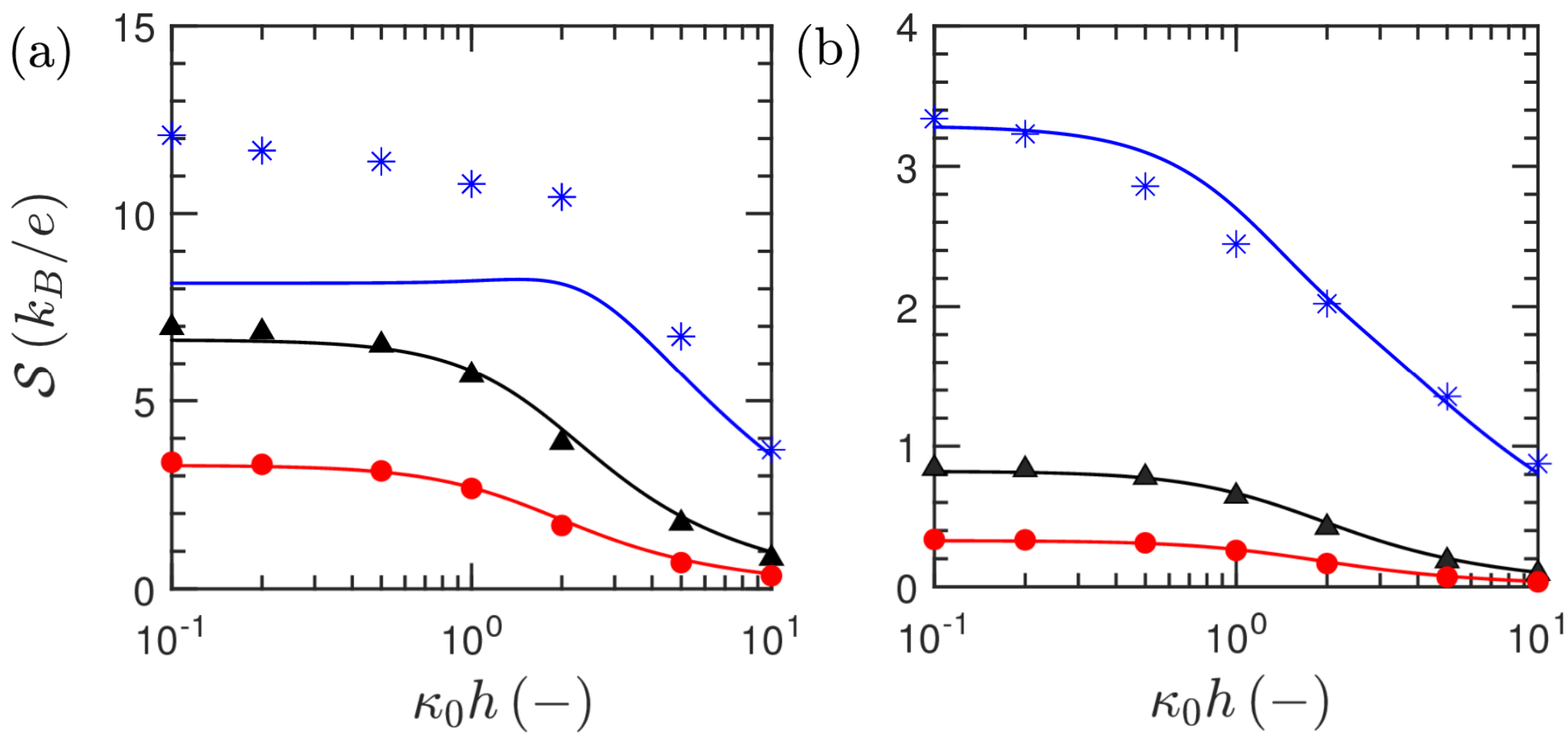

**FIG. 3.** Seebeck coefficient as a function of the non-dimensional channel width ($\kappa_0 h$) for (a) the weak electrolyte solution picric acid in 2-methoxyethanol and (b) a fully dissociated electrolyte. Results of full numerical simulations (symbols) based on the extended PNP equations are compared with the analytical solution (lines) of Eq. (4). The red circles (and the corresponding solid line) correspond to $\zeta = -10\,\text{mV}$, the black triangles (and the corresponding line) represent $\zeta = -25\,\text{mV}$, and the blue stars (and the corresponding line) stand for $\zeta = -100$ mV. The temperature was set to $T_0 = 343\,\text{K}$, and the temperature difference to $\Delta T = 10\,\text{K}$. The parameters for picric acid in 2-methoxyethanol were obtained from [18].

Figure 3a shows $\mathcal{S}$ as a function of the non-dimensional channel width $\kappa_0 h$ and for different values of the $\zeta$ potential for picric acid in 2-methoxyethanol. In that case, the dissociation coefficient is given by the relationship $\mathcal{K}_{eq} = \mathcal{C} \exp\left(\alpha_0 + \alpha_1 T + \alpha_2/T + \alpha_3 \ln T\right)$, where $\mathcal{C} = 1\,\text{M}$ and $\alpha_i\ (i = 0-3)$ are constants [18]. For this system, no solute diffusion coefficients appear to be available from the literature, which is why a common value of $10^{-9}\ \text{m}^2/\text{s}$ was assumed for all solutes. Considering that $\mathcal{S}$ from Eq. (4) is a local quantity, for channel of finite length the thermovoltage is obtained via $\Delta V = (dT/dx)\int \mathcal{S}\,dx$, where the integral is along the entire channel. Here, the results are computed for sufficiently small $\Delta T$ $\left(\Delta T/T_0 \ll 1\right)$, which allows approximating the integral by $\Delta\varphi \approx \Delta T\,\mathcal{S}\big|_{T=T_0}$. For a comparison between weak and fully dissociated electrolytes, the Seebeck coefficient obtained for the latter case is shown in Figure

3b. Furthermore, for both sets of data, the numerically computed $\mathcal{S}$ is compared with the analytical results. This provides a validation of the analytical model and an assessment of its range of validity. The numerical simulations are carried out by prescribing the corresponding surface charge density $q$ for each $\zeta$ potential value considered. A mapping between $q$ and $\zeta$ is achieved based on the non-linear Guoy-Chapman model for non-overlapping EDLs [23]. This mapping is reasonably accurate for the considered range of $\kappa_0 h$ values. At smaller values of $\zeta$, a good agreement between the analytical and the numerical results is found, supporting the robustness of our analysis. The deviation that can be noticed between both approaches at higher $\zeta$ values is due to the Debye-Hückel approximation, which is limited to the values below $|ev\zeta/k_B T| \approx 1$. As indicated above, the induced thermoelectric potential is at its maximum in the regime of large EDL overlap (i.e., for $\kappa_0 h \to 0$) and tends towards zero as $\kappa_0 h \to \infty$. Comparing Figure 3a and 3b, one can find that the thermoelectric response of the weak electrolyte is 4-10 times higher than that of the fully dissociated electrolyte, depending on the $\zeta$ potential value.

In conclusion, our analysis reveals an enormous potential of confined weak electrolytes for thermoelectric energy conversion. The main driver for the thermoelectric response is the counter-ion concentration gradient that forms in a nanochannel due to the temperature-dependent dissociation of neutral solute into ions. For the specific weak electrolyte picric acid in 2-methoxyethanol, we predict an up to tenfold increase of the Seebeck coefficient compared to a fully dissociated electrolyte solution. Taking into consideration the broad spectrum of weak electrolytes, characterized by the parameter space $\{\mathcal{K}_{eq}, d\mathcal{K}_{eq}/dT\}$, in the future it should be possible to identify specific materials that yield a giant Seebeck coefficient, potentially exceeding all values that have been reported in the literature up to now.

Our analysis also significantly contributes to developing a theoretical framework for transport processes in electrolytes under non-isothermal conditions. The extended Nernst-Planck equations we have developed should be applicable to many non-isothermal scenarios in which weak electrolytes are present, provided that the assumption of a fast dissociation-association kinetics holds. Our model is minimal in the sense that a symmetric binary electrolyte is considered, together with the un-dissociated form of the weak electrolyte. In the future, the model can be extended to account for asymmetric electrolytes, multiple species, and Stern layer effects, to name a few.

## Data availability

The datasets generated during and/or analyzed during the current study are available in the TRANSLATE Zenodo repository: https://zenodo.org/communities/translateh2020/

## Acknowledgement

This project has received funding from the European Union's Horizon 2020 research and innovation program under grant agreement No 964251.

## Disclaimer

This publication reflects only the author's view and the European Commission is not responsible for any use that may be made of the information it contains.

[1] C. Forman, I. K. Muritala, R. Pardemann, and B. Meyer, Estimating the global waste heat potential, Renewable and Sustainable Energy Reviews **57**, 1568 (2016).

[2] E. Garofalo, M. Bevione, L. Cecchini, F. Mattiussi, and A. Chiolerio, Waste Heat to Power: Technologies, Current Applications, and Future Potential, Energy Technology **8**, 2000413 (2020).

[3] S. Sun, M. Li, X.-L. Shi, and Z.-G. Chen, Advances in Ionic Thermoelectrics: From Materials to Devices, Advanced Energy Materials **13**, 2203692 (2023).

[4] Y.-H. Pai, J. Tang, Y. Zhao, and Z. Liang, Ionic Organic Thermoelectrics with Impressively High Thermopower for Sensitive Heat Harvesting Scenarios, Advanced Energy Materials **13**, 2202507 (2023).

[5] C. Zhang, X.-L. Shi, Q. Liu, and Z.-G. Chen, Hydrogel-Based Functional Materials for Thermoelectric Applications: Progress and Perspectives, Advanced Functional Materials **34**, 2410127 (2024).

[6] S. Jia, W. Qian, P. Yu, K. Li, M. Li, J. Lan, Y.-H. Lin, and X. Yang, Ionic thermoelectric materials: Innovations and challenges, Materials Today Physics **42**, 101375 (2024).

[7] H. Cheng, Z. Wang, Z. Guo, J. Lou, W. Han, J. Rao, and F. Peng, Cellulose-based thermoelectric composites: A review on mechanism, strategies and applications, International Journal of Biological Macromolecules **275**, 132908 (2024).

[8] Z. A. Akbar, Y. T. Malik, D.-H. Kim, S. Cho, S.-Y. Jang, and J.-W. Jeon, Self-Healable and Stretchable Ionic-Liquid-Based Thermoelectric Composites with High Ionic Seebeck Coefficient, Small **18**, 2106937 (2022).

[9] Z. Liu, H. Cheng, H. He, J. Li, and J. Ouyang, Significant Enhancement in the Thermoelectric Properties of Ionogels through Solid Network Engineering, Advanced Functional Materials **32**, 2109772 (2022).

[10] B. Kim, J. Na, H. Lim, Y. Kim, J. Kim, and E. Kim, Robust High Thermoelectric Harvesting Under a Self-Humidifying Bilayer of Metal Organic Framework and Hydrogel Layer, Advanced Functional Materials **29**, 1807549 (2019).

[11] W. Zhao, Y. Zheng, M. Jiang, T. Sun, A. Huang, L. Wang, W. Jiang, and Q. Zhang, Exceptional n-type thermoelectric ionogels enabled by metal coordination and ion-selective association, Science Advances **9**, eadk2098 (2023).

[12] M. Dietzel and S. Hardt, Thermoelectricity in Confined Liquid Electrolytes, Phys. Rev. Lett. **116**, 225901 (2016).
[13] R. Sarma and S. Hardt, Giant Thermoelectric Response of Confined Electrolytes with Thermally Activated Charge Carrier Generation, Phys. Rev. Lett. **132**, 098001 (2024).
[14] M. A. Gebbie, M. Valtiner, X. Banquy, E. T. Fox, W. A. Henderson, and J. N. Israelachvili, Ionic liquids behave as dilute electrolyte solutions, Proceedings of the National Academy of Sciences **110**, 9674 (2013).
[15] M. A. Gebbie, H. A. Dobbs, M. Valtiner, and J. N. Israelachvili, Long-range electrostatic screening in ionic liquids, Proceedings of the National Academy of Sciences **112**, 7432 (2015).
[16] L. Fu, L. Joly, and S. Merabia, Giant Thermoelectric Response of Nanofluidic Systems Driven by Water Excess Enthalpy, Phys. Rev. Lett. **123**, 138001 (2019).
[17] A. Würger, Thermoelectric Ratchet Effect for Charge Carriers with Hopping Dynamics, Phys. Rev. Lett. **126**, 068001 (2021).
[18] G. Franchini, E. Ori, C. Preti, L. Tassi, and G. Tosi, A conductometric study of dissociation of picric acid in 2-methoxyethanol and 1,2-ethanediol from −10 to 80 °C, Can. J. Chem. **65**, 722 (1987).
[19] G. C. Franchini, L. Tassi, and G. Tosi, The ethane-1,2-diol–2-methoxyethanol solvent system, J. Chem. Soc., Faraday Trans. 1 **83**, 3129 (1987).
[20] G. Franchini, A. Marchetti, C. Preti, L. Tassi, and G. Tosi, An Approach to the Problem of the Dependence of the Dissociation Constant of Weak Electrolytes on the Temperature and on the Solvent Composition in the Ethane-1,2-dio1-2-Methoxyethanol Solvent System, 1 **85**, 1697 (1989).
[21] M. Paabo, R. G. Bates, and R. A. Robinson, Dissociation of Acetic Acid-eZ3 in Aqueous Solution and Related Isotope Effects from 0 to 50°, The Journal of Physical Chemistry **70**, 540 (1966).
[22] H. S. Harned and N. D. Embree, The Ionization Constant of Formic Acid from 0 to 60°1, J. Am. Chem. Soc. **56**, 1042 (1934).
[23] See Supplemental Material for derivation details.
[24] M. J. Blandamer, J. Burgess, P. P. Duce, R. E. Robertson, and J. W. M. Scott, The dependence of acid dissociation constants in water on temperature as expressed by the Gurney equation, Can. J. Chem. **59**, 2845 (1981).
[25] H. S. Harned and R. W. Ehlers, The Dissociation Constant of Acetic Acid from 0 to 60° Centigrade, J. Am. Chem. Soc. **55**, 652 (1933).
[26] H. S. Harned and N. D. Embree, The Ionization Constant of Formic Acid from 0 to 60°1, J. Am. Chem. Soc. **56**, 1042 (1934).
[27] R. R. Schroeder and I. Shain, Application of the potentiostatic method. Determination of the rate constant for the dissociation of acetic acid, J. Phys. Chem. **73**, 197 (1969).
[28] P. Delahay and W. Vielstich, Kinetics of the Dissociation of Weak Acids and Bases—Application of Polarography and Voltammetry at Constant Current, J. Am. Chem. Soc. **77**, 4955 (1955).

## Supplemental Information for "Strong thermoelectric response of nanoconfined weak electrolytes"

### §1: Derivation of the Nernst-Planck equation for weak electrolytes

The following derivation refers to a weak electrolyte that either exists in the form of a neutral solute or dissociated into an anion or cation. The Nernst-Planck equation (NPE) describing the evolution of the concentration $c_k$ of each species *k*, therefore, reads

$$\frac{\partial c_k}{\partial t}+\mathbf{u}\cdot\nabla c_k+\nabla\cdot\mathbf{j}_k=\mathcal{G}_k\,, \tag{S1}$$

where $k=n$ for the neutral solute, $k=+$ for the cation, and $k=-$ for the anion. The source term $\mathcal{G}_k$ (defined in the main text) describes the rate of generation of the $k^{\text{th}}$ species per unit volume. Furthermore, $\mathbf{u}$ represents the flow velocity vector, $-\mathbf{j}_k=D_k\nabla c_k+e\nu_k\mu_k c_k\nabla\phi$, with $D_k$, $\mu_k$, $\nu_k$ being the Fickian diffusion coefficient, electrophoretic mobility, and ionic valence of each species, respectively, while $\phi$ is the total electric potential. As mentioned in the main text, we consider here a symmetric weak electrolyte solution. Therefore, $D_k\equiv D$, $\mu_k\equiv\mu\left(=D/k_BT\right)$ and $\nu_+=-\nu_-=\nu$. The flow within the channel is considered to be fully-developed, implying $\mathbf{u}=\left(u,0\right)$. Defining the non-dimensional variables $\bar{x}=x/l$, $\bar{z}=z/h$, $\bar{t}=tU/l$, $\bar{u}=u/U$, $\bar{\nu}_k=\nu_k/\nu$, $\bar{c}_k=c_k/c_r$, $\bar{\phi}=e\nu\phi/k_BT_r$, $\theta=\left(T-T\right)_r/\Delta T$, $\hat{\theta}+1=T/T_r$, one can cast Eq. (S1) into the following non-dimensional form

$$\begin{aligned}&A^2\mathrm{Pe}\left[\frac{\partial\bar{c}_k}{\partial\bar{t}}+\bar{u}\frac{\partial\bar{c}_k}{\partial\bar{x}}\right]-A^2\frac{1}{D}\frac{\partial}{\partial\bar{x}}\left[D\frac{\partial\bar{c}_k}{\partial\bar{x}}+\bar{\nu}_k\frac{D}{1+\hat{\theta}}\bar{c}_k\frac{\partial\bar{\phi}}{\partial\bar{x}}\right]\\&\qquad-\frac{1}{D}\frac{\partial}{\partial\bar{z}}\left[D\frac{\partial\bar{c}_k}{\partial\bar{z}}+\bar{\nu}_k\frac{D}{1+\hat{\theta}}\bar{c}_k\frac{\partial\bar{\phi}}{\partial\bar{z}}\right]=A^2\delta_k\mathrm{Da}\left(\bar{c}_n-\bar{\mathcal{K}}^{-1}\bar{c}_+\bar{c}_-\right)\end{aligned}\;. \tag{S2}$$

The characteristic velocity $U\left(=u_{\mathrm{HS}}\right)$ is the (thermally induced) Helmholtz-Smoluchowski velocity, the magnitude of which will be computed below. The reference concentration is $c_r = c_b$ (defined in the main text), $\nu$ is the valence of the cationic species, while the reference temperature $T_r$ is the mean temperature of the channel. Furthermore, $\bar{\mathcal{K}} = \mathcal{K}_d / \mathcal{K}_a c_r$, $\mathrm{Pe} = Ul/D$ is the ionic Péclet number, and $\mathrm{Da} = l^2 \mathcal{K}_d / D$ is the Damköhler number. An estimation of the magnitude of $\mathrm{Da}$ is essential to determine how fast the dissociation process attains its equilibrium relative to diffusion. For weak electrolytes, usually $\mathcal{K}_d \approx O\left(10^5 - 10^6\right)$ 1/s [S1,S2]. Considering $D \approx O\left(10^{-9} - 10^{-12}\right)$ m$^2$/s, one finds $\mathrm{Da} \approx O\left(10^2 - 10^6\right) \gg 1$. This indicates a quasi-instantaneous local equilibrium between the dissociated and undissociated species for a weak electrolyte solution.

To assess the importance of advection for the species distribution, we first need to estimate the characteristic velocity $U$ and the order of magnitude of $\mathrm{Pe}$. This necessitates solving the steady-state Navier-Stokes equation (NSE) for a fully-developed flow configuration. With $\boldsymbol{\tau}_V$ denoting the viscous stress tensor and $\boldsymbol{\tau}_M$ the Maxwell stress tensor, the corresponding NSE is given by

$$\nabla \cdot \boldsymbol{\tau}_V + \nabla \cdot \boldsymbol{\tau}_M = 0, \tag{S3}$$

where $\boldsymbol{\tau}_V = \eta\left[\nabla \mathbf{u} + \left(\nabla \mathbf{u}\right)^{\mathrm{T}}\right]$ and $\boldsymbol{\tau}_M = \epsilon\left[\nabla\phi\nabla\phi - \left(\nabla\phi\right)^2 \mathbf{I}/2\right]$. The quantities $\eta$, $\epsilon$ and $\mathbf{I}$ in Eq. (S3) represent the dynamic viscosity, dielectric permittivity and unit tensor, respectively. A combined electroosmotic and thermoosmotic flow results from the induced thermoelectric field $E$, the temperature dependence of $D/\mu$, and the temperature-dependent dielectric permittivity. For $A^2 \ll 1$, Eq. (S3) can be simplified to

$$\frac{\partial}{\partial z}\left(\eta \frac{\partial u}{\partial z}\right) = \epsilon E \frac{\partial^2 \psi}{\partial z^2} + \frac{1}{2}\frac{\partial T}{\partial x}\left[\epsilon_T \left(\frac{\partial \psi}{\partial z}\right)^2 - \frac{2e^2\nu^2 c_0}{k_B T_0^2}\left(\psi^2 - \psi_{\mathrm{c}}^2\right) - \frac{4e^2\nu^2 c_0}{k_B T}\psi_{\mathrm{c}}\frac{\partial \psi_{\mathrm{c}}}{\partial T}\right], \tag{S4}$$

where $c_0 = \sqrt{\mathcal{K}_{eq} c_n}$, $\epsilon_T = d\epsilon/dT$, and $\psi_c$ is the electric potential at $z = 0$. Inserting the derivatives of $\psi$ (evaluated under the Debye-Hückel approximation) into Eq. (S4) and solving for $u$ using the no-slip boundary condition at the walls, we obtain

$$\begin{aligned} u = &-\frac{\epsilon\zeta^2}{\eta}\frac{1}{8\cosh^2(\kappa h)}\frac{1}{T_0}\frac{\partial T}{\partial x}\left\{\left[\frac{\cosh(2\kappa z)-\cosh(2\kappa h)}{2}+\kappa^2 h^2\left(1-\frac{z^2}{h^2}\right)\right]\left(\frac{1}{1+\hat{\theta}}-MT_0\right)\right. \\ &\left.-2\kappa^3 h^3\left(1-\frac{z^2}{h^2}\right)\tanh(\kappa h)\left(\frac{1}{1+\hat{\theta}}+MT_0\right)\right\}+\frac{\epsilon\zeta}{\eta}\left[\frac{\cosh(\kappa z)}{\cosh(\kappa h)}-1\right]E \end{aligned}, \quad \text{(S5)}$$

where $M = \epsilon_T/\epsilon$.

Next, we substitute $E$ derived in Eq. (4) into Eq. (S5). Abbreviating $\epsilon\zeta^2(\partial T/\partial x)/\eta T_0$ by $u_{\mathrm{HS}}$, the local velocity becomes

$$\begin{aligned} \frac{u}{u_{\mathrm{HS}}} = &T_0\left[-\frac{1}{\mathcal{K}_{eq}\left(2+\sqrt{\mathcal{K}_{eq}/c_n}\right)}\frac{d\mathcal{K}_{eq}}{dT}+\frac{1}{T}\right]\frac{\tanh(\kappa h)}{\kappa h}\left[\frac{\cosh(\kappa z)}{\cosh(\kappa h)}-1\right] \\ &+\frac{1}{8\cosh^2(\kappa h)}\left\{\left[\frac{\cosh(2\kappa z)-\cosh(2\kappa h)}{2}+\kappa^2 h^2\left(1-\frac{z^2}{h^2}\right)\right]\left(MT_0-\frac{1}{1+\hat{\theta}}\right)\right. \\ &\left.+2\kappa^3 h^3\left(1-\frac{z^2}{h^2}\right)\tanh(\kappa h)\left(MT_0+\frac{1}{1+\hat{\theta}}\right)\right\} \end{aligned}. \quad \text{(S6)}$$

For symmetric boundary conditions at the walls, the maximum velocity is attained at $z = 0$, and is given by

$$\begin{aligned} \frac{u_{\max}}{u_{\mathrm{HS}}} = &T_0\left[-\frac{1}{\mathcal{K}_{eq}\left(2+\sqrt{\mathcal{K}_{eq}/c_n}\right)}\frac{d\mathcal{K}_{eq}}{dT}+\frac{1}{T}\right]\frac{\tanh(\kappa h)}{\kappa h}\left[\frac{1}{\cosh(\kappa h)}-1\right] \\ &+\frac{1}{8}\left[\left(\frac{\kappa^2 h^2}{\cosh^2(\kappa h)}-\tanh^2(\kappa h)\right)\left(MT_0-\frac{1}{1+\hat{\theta}}\right)+2\kappa^3 h^3\frac{\tanh(\kappa h)}{\cosh^2 \kappa h}\left(MT_0+\frac{1}{1+\hat{\theta}}\right)\right] \end{aligned}. \quad \text{(S7)}$$

For a quantitative evaluation of $u_{\max}$, we consider the weak electrolyte solution picric acid in 2-methoxyethanol, with $T = T_r = 313\,\mathrm{K}$ and $c_b\ (= c_n + c_i)$ of 0.1 M. In this case, $\mathcal{K}_{eq} \approx O(10^{-5})$ M and $d\mathcal{K}_{eq}/dT \approx O(-10^{-6})$ M/K, $\epsilon \approx O(10)\times 8.854\times 10^{-12}$ F/m, $\epsilon_T \approx O(10^{-2})\,\mathrm{K}^{-1}$, $D \approx O\left(10^{-9}\right)$ m$^2$/s, $\eta \approx O\left(10^{-2}\right)$ Pa·s [S3]. With these parameters, the maximum flow velocity is estimated to be $u = 5.48u_{\mathrm{HS}}$ at $\kappa h = 1.75$. Considering $\Delta T \approx O(10)$ K, $\zeta \approx O(10)$ mV and $l \approx O\left(10^{-6}\right)$ m, the Péclet number is limited to $\mathrm{Pe} \lesssim O\left(10^{-2}\right)$. This suggests that the advective term in Eq. (S2) can be neglected in the further analysis.

In dimensional form, Eq. (S2), therefore, simplifies to

$$-\frac{\partial}{\partial x}\left(D\frac{\partial c_k}{\partial x}+e\nu_k\mu c_k\frac{\partial\phi}{\partial x}\right)-\frac{\partial}{\partial z}\left(D\frac{\partial c_k}{\partial z}+e\nu_k\mu c_k\frac{\partial\phi}{\partial x}\right)=\delta_k\left(\mathcal{K}_d c_n-\mathcal{K}_a c_+ c_-\right).$$

(S8)

The stoichiometric coefficient $\delta_k$ takes the value -1 when $k$ refers to a neutral solute, and 1 for the charged species. For a symmetric electrolyte, the reaction term in Eq. (S8) can be eliminated after some algebraic manipulation. This includes multiplying Eq. (S8) by $\delta_k = 1$ for $k = n$ and adding up the resulting equation and the corresponding equations for $k = (+,-)$. Finally one obtains

$$\nabla\cdot\left(D\nabla c_n + D\nabla c_i + e\nu_i\mu c_i\nabla\phi\right)=0, \tag{S9}$$

where $i = (+,-)$. In the derivation of Eq. (S9), local chemical equilibrium was assumed, based on the large Damköhler number, as discussed above. This allows reducing the number of equations to two, while $c_n$ is obtained from

$$\mathcal{K}_{eq}=\frac{c_+c_-}{c_n}, \tag{S10}$$

Equations (S9) and (S10) together describe the transport of solutes in a weak electrolyte solution.

### §2: Distribution of charged species within the nanochannel

In terms of the leading-order contribution in $A$, one obtains from Eq. (S9)

$$\frac{\partial}{\partial z}\left[D\frac{\partial c_n}{\partial z}\right]+\frac{\partial}{\partial z}\left[D\frac{\partial c_i}{\partial z}+e\nu_i\mu c_i\frac{\partial\phi}{\partial z}\right]=0, \qquad \text{(S11)}$$

where $\mu = D/k_B T$ and $\partial\phi/\partial z = \partial\psi/\partial z$. Integrating Eq. (S11) with respect to $z$, subject to a symmetry condition along the channel centerline, we arrive at

$$D\frac{\partial c_n}{\partial z}+D\left(\frac{\partial c_i}{\partial z}+\frac{e\nu_i}{k_B T}c_i\frac{\partial\psi}{\partial z}\right)=0\,. \qquad \text{(S12)}$$

In the electroneutral region of the symmetric weak electrolyte solution considered herein, $c_i=\sqrt{c_n\mathcal{K}_{eq}}=c_0\ \ (i=+,-)$, where $c_0$ does not depend on $z$. A Taylor's series expansion around the point of electroneutrality for each species gives

$$c_+\approx c_0+\varepsilon\tilde{c}_+ +\ldots, \qquad \text{(S13a)}$$

$$c_-\approx c_0+\varepsilon\tilde{c}_- +\ldots, \qquad \text{(S13b)}$$

$$c_n\approx\frac{1}{\mathcal{K}_{eq}}\left[c_0^2+\varepsilon c_0\left(\tilde{c}_+ +\tilde{c}_-\right)+\ldots\right], \qquad \text{(S13c)}$$

where $\varepsilon$ is a small parameter, which is related to the $\zeta$ potential of the wall material. The higher the $\zeta$ potential (and, correspondingly, the surface charge density), the larger will be $\varepsilon$. Therefore, the assumption of small $\varepsilon$ aligns with the Debye-Hückel approximation. Simultaneously, the expansion of $\psi$ results in $\psi\approx\psi_0+\varepsilon\tilde{\psi}+\ldots$To leading order in $\varepsilon$, Eq. (S12) reduces to $c_0\,\partial\psi_0/\partial z=0$. This implies $\psi_0=const.$, and without loss of generality we set $\psi_0=0$.

To $O(\varepsilon)$, Eq. (S12) gives

$$\frac{c_0}{\mathcal{K}_{eq}}\frac{\partial}{\partial z}\left(\tilde{c}_+ +\tilde{c}_-\right)+\left(\frac{\partial\tilde{c}_+}{\partial z}+\frac{e\nu_+}{k_B T}c_0\frac{\partial\tilde{\psi}}{\partial z}\right)=0, \qquad \text{(S14a)}$$

$$\frac{c_0}{\mathcal{K}_{eq}}\frac{\partial}{\partial z}\left(\tilde{c}_+ +\tilde{c}_-\right)+\left(\frac{\partial\tilde{c}_-}{\partial z}+\frac{e\nu_-}{k_B T}c_0\frac{\partial\tilde{\psi}}{\partial z}\right)=0\,. \qquad \text{(S14b)}$$

A little algebraic manipulations of (S14a,b) for $\nu_+ = -\nu_- = \nu$ thus yields

$$\frac{\partial}{\partial z}\left[\tilde{c}_+ - \tilde{c}_- + \frac{2e\nu}{k_B T} c_0 \tilde{\psi}\right] = 0 , \tag{S15a}$$

$$\left(2\frac{c_0}{\mathcal{K}_{eq}} + 1\right)\frac{\partial}{\partial z}\left(\tilde{c}_+ + \tilde{c}_-\right) = 0 . \tag{S15b}$$

The solution of Eq. (S15a) reads $\tilde{c}_+ - \tilde{c}_- + \left(2e\nu / k_B T\right) c_0 \tilde{\psi} = const. = \mathcal{C}_1$, while solving Eq. (S15b), one finds $\tilde{c}_+ + \tilde{c}_- = const. = \mathcal{C}_2$. To compute the constant $\mathcal{C}_1$, we apply the condition $\tilde{\psi} = 0$ for $\tilde{c}_+ - \tilde{c}_- = 0$, which gives $\mathcal{C}_1 = 0$. To compute $\mathcal{C}_2$, we add Eqs. (S13a) and (S13b) to $O(\varepsilon)$ and employ the condition $c_+ + c_- = 2c_0$ in the electroneutral region, resulting in $\mathcal{C}_2 = 0$. Now, since $\tilde{c}_+ + \tilde{c}_- = 0$ and $c_0$ is $z$-independent, it is clear from Eq. (S13c) that $c_n$ is $z$-independent as well (since $T$ is $z$-independent, as shown in §4). Thus, integrating Eq. (S12) in $z$ -direction, one obtains

$$c_i = \gamma \exp\left(-\frac{e\nu_i \psi}{k_B T}\right), \tag{S16}$$

where $\gamma$ is an integration constant. To evaluate $\gamma$, we set $\psi = 0$ in Eq. (S16), which gives $\gamma = c_i = c_0$, the ion concentration in the electroneutral region. Note that, since $\mathcal{K}_{eq} = \mathcal{K}_{eq}\left(T\right)$, $c_0$ is not a constant but varies axially in the presence of an axial temperature gradient. Substituting $\gamma$ into Eq. (S16), we find the distribution of the charged species in a weak electrolyte solution.

**§3: Electric potential distribution**

To first order in $A$, the Poisson equation introduced in the main text can be approximated by

$$\epsilon \frac{\partial^2 \psi}{\partial z^2} = -\rho_f . \tag{S17}$$

Despite $\epsilon = \epsilon(T)$, Eq. (S17) shows that to leading order in $A$, the temperature dependence of the dielectric permittivity has no influence on the electric potential distribution, since the temperature only varies in axial direction. Inserting $\rho_f = e\sum_i \nu_i c_i$ in Eq. (S17), one obtains

$$\epsilon \frac{d^2\psi}{dz^2} = -e\sum_i \nu_i c_i \, . \tag{S18}$$

Substituting $c_i$ given by Eq. (S16) into Eq. (S18) yields for a symmetric electrolyte

$$\epsilon \frac{d^2\psi}{dz^2} = 2e\nu c_0 \sinh\left(\frac{e\nu\psi}{k_B T}\right). \tag{S19}$$

Here, $c_0$ and $c_n$ are related through $c_0 = \sqrt{\mathcal{K}_{eq} c_n}$. For small $\zeta$ potential $\left(\left|e\nu\zeta / k_B T\right| < 1\right)$, the right-hand side (RHS) of Eq. (S19) can be linearized using the Debye–Hückel approximation, yielding

$$\frac{d^2\psi}{dz^2} = \kappa^2 \psi \, . \tag{S20}$$

In Eq. (S20), $\kappa = \sqrt{2e^2 \nu^2 c_0 / \epsilon k_B T}$ denotes the local Debye parameter. With the boundary condition $\psi = \zeta$ at the walls, the solution of Eq. (S20) reads

$$\psi = \zeta \frac{\cosh(\kappa z)}{\cosh(\kappa h)}. \tag{S21}$$

**§4: Temperature distribution**

The temperature distribution within the nanochannel can be computed by solving the energy equation. Including viscous dissipation and Joule heating, the energy equation reads

$$\rho c_p \left[\frac{\partial T}{\partial t} + \mathbf{u} \cdot \nabla T\right] = \nabla \cdot \left(\lambda \nabla T\right) + \dot{q}_\eta + \dot{q}_\phi \, , \tag{S22}$$

where $\rho$, $c_p$ and $\lambda$ denote the density, specific heat capacity and thermal conductivity of the fluid, respectively. For a fully developed flow, the heat generation (per unit volume) due to viscous dissipation is given by $\dot{q}_\eta = \eta\left(\partial u/\partial z\right)^2$. The Joule heating is caused by the passage of an electric current $I$ through the electrolyte solution. To determine the Seebeck coefficient, we study a scenario with vanishing current. Therefore, Joule heating does not play a role in the present context. Introducing the non-dimensional variables defined in §1 in Eq. (S22), one can write

$$A^2\left[\mathrm{Pe}_T\left(\frac{\partial\theta}{\partial\bar{t}}+\bar{u}\frac{\partial\theta}{\partial\bar{x}}\right)-\frac{1}{\lambda}\frac{\partial}{\partial\bar{x}}\left(\lambda\frac{\partial\theta}{\partial\bar{x}}\right)\right]=\frac{1}{\lambda}\frac{\partial}{\partial\bar{z}}\left(\lambda\frac{\partial\theta}{\partial\bar{z}}\right)+\frac{\eta U^2}{\lambda\Delta T}\left(\frac{\partial\bar{u}}{\partial\bar{z}}\right)^2, \tag{S23}$$

where $\mathrm{Pe}_T = \rho U c_p h/\lambda$ is the thermal Péclet number.

To estimate the contributions of advection and viscous dissipation on the temperature distribution, we need to find the order of magnitude of these terms. The magnitude of viscous dissipation (taken into account by the second term on the RHS of Eq. (S23)) can be assessed by substituting $U = 5.48u_{\mathrm{HS}}$ (the maximum flow velocity derived in §1). With $\max\left\{\left(\partial\bar{u}/\partial\bar{z}\right)^2\right\}=1$, $\lambda \approx O\left(10^{-1}\right)$W/m·K, $\Delta T \approx O\left(10\right)$K and considering the values of the remaining parameters as mentioned in §1, the viscous dissipation is estimated to be

$$\frac{\eta U^2}{\lambda\Delta T}\left(\frac{\partial\bar{u}}{\partial\bar{z}}\right)^2 \lesssim \frac{\eta}{\lambda\Delta T}\left(5.48\frac{\epsilon\zeta^2}{\eta T_r}\frac{\Delta T}{l}\right)^2 \approx O\left(10^{-17}\right). \tag{S24}$$

This suggests that the effect of viscous dissipation can be safely ignored.

To determine the role of advection on the temperature field, we need to estimate the magnitude of $\mathrm{Pe}_T$. With $U = 5.48u_{\mathrm{HS}}$, one obtains

$$\mathrm{Pe}_T \approx 5.48A\frac{\rho c_p}{\lambda}\frac{\epsilon\zeta^2}{\eta T_0}\Delta T \approx O\left(10^{-7}A\right). \tag{S25}$$

Here, $\rho \approx O\left(10^3\right)$kg/m$^3$ and $c_p \approx O\left(10^3\right)$J/kg·K. A negligibly small $\mathrm{Pe}_T$ indicates that diffusive heat transport dominates over advective transport. Therefore, in terms of the leading-order contribution in $A$, Eq. (S23) simplifies to

$$\frac{1}{\lambda}\frac{\partial}{\partial \bar{z}}\left(\lambda\frac{\partial\theta}{\partial \bar{z}}\right)=0\,. \tag{S26}$$

Solving Eq. (S26) assuming a vanishing heat flux across the walls at $z=\pm h$, one finds $T=const.$ in the *z*-direction, i.e., $T=T\left(x\right)$. With $T_0$ and $T_e\left(=T_0+\Delta T\right)$ denoting the cold and hot reservoir temperatures, respectively, the temperature distribution within the nanochannel is given by $T(x)=T_0+\left(x/l\right)\Delta T$.

## §5: Derivation of the Seebeck coefficient

The electric current of the considered weak electrolyte solution is given by

$$\mathbf{I}=e\nu D\int_0^h\left[\left(\nabla c_+-\nabla c_-\right)+\frac{e\nu}{k_BT}\left(c_++c_-\right)\left(\nabla\psi-\mathbf{E}\right)\right]dz\,. \tag{S27}$$

The induced electric field *E*, and accordingly the Seebeck coefficient $\mathcal{S}\left(=E/(dT/dx)\right)$ can be computed by setting $I=0$. Substituting $c_i$ given by Eq. (S16) into Eq. (S27), one finds

$$\begin{aligned}&-\frac{e\nu}{k_BT}E\int_0^h c_0\left[\exp\left(-\frac{e\nu\psi}{k_BT}\right)+\exp\left(\frac{e\nu\psi}{k_BT}\right)\right]dz\\&=\int_0^h\frac{dc_0}{dx}\left[\exp\left(\frac{e\nu\psi}{k_BT}\right)-\exp\left(-\frac{e\nu\psi}{k_BT}\right)\right]dz-\frac{e\nu}{k_BT^2}\frac{dT}{dx}\int_0^h\psi c_0\left[\exp\left(\frac{e\nu\psi}{k_BT}\right)+\exp\left(-\frac{e\nu\psi}{k_BT}\right)\right]dz\end{aligned}\,. \tag{S28}$$

Inserting $c_0=\sqrt{c_n\mathcal{K}_{eq}}$ into Eq. (S28) and some algebra yields

$$\begin{aligned}-\frac{e\nu}{k_BT}\frac{E}{dT/dx}\int_0^h\left[\exp\left(-\frac{e\nu\psi}{k_BT}\right)+\exp\left(\frac{e\nu\psi}{k_BT}\right)\right]dz&=\int_0^h\left[\frac{1}{2\mathcal{K}_{eq}}\frac{d\mathcal{K}_{eq}}{dT}\left\{\exp\left(\frac{e\nu\psi}{k_BT}\right)-\exp\left(-\frac{e\nu\psi}{k_BT}\right)\right\}\right]dz\\&+\int_0^h\left[\frac{1}{2c_n}\frac{dc_n}{dT}\left\{\exp\left(\frac{e\nu\psi}{k_BT}\right)-\exp\left(-\frac{e\nu\psi}{k_BT}\right)\right\}-\frac{e\nu\psi}{k_BT^2}\left\{\exp\left(\frac{e\nu\psi}{k_BT}\right)+\exp\left(-\frac{e\nu\psi}{k_BT}\right)\right\}\right]dz.\end{aligned}$$

(S29)

Further simplifying Eq. (S29), we obtain

$$\mathcal{S} = \frac{-\frac{k_B T}{2e\nu}\left(\frac{1}{c_n}\frac{dc_n}{dT}+\frac{1}{\mathcal{K}_{eq}}\frac{d\mathcal{K}_{eq}}{dT}\right)\int_0^h \sinh\left(\frac{e\nu\psi}{k_B T}\right)dz+\frac{1}{T}\int_0^h \psi\cosh\left(\frac{e\nu\psi}{k_B T}\right)dz}{\int_0^h \cosh\left(\frac{e\nu\psi}{k_B T}\right)dz}. \quad \text{(S30)}$$

Evaluation of the integrals under the Debye–Hückel approximation with series expansion of the hyperbolic functions gives Eq. (4) of the main text.

## §6: Computational approach for evaluating the Seebeck coefficient

We perform finite-element simulations to numerically compute the Seebeck coefficient for weak electrolytes. This includes solving the modified Nernst-Planck equations (Eqs. (S9), (S10)) and the Poisson equation $\nabla\cdot\left(\epsilon\nabla\phi\right)=-e\sum_k \nu_k c_k$ together with the energy equation $\nabla^2 T=0$. The computations are carried out in two dimensions within the domain sketched in Fig. S1.

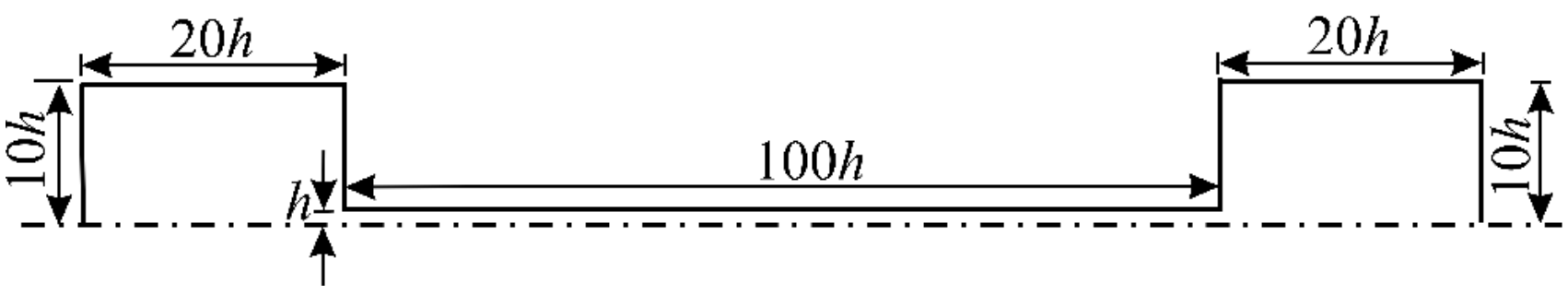

**FIGURE S1:** Schematic drawing of the two-dimensional computational domain comprising a parallel-plate channel of width $2h$ between two reservoirs.

The boundary conditions comprise the impermeability for species mass flux $(j_{k,z}=0)$ and thermal insulation $(\partial T/\partial z|_{z=\pm h}=0)$ for the channel walls as well as the reservoirs. This, however, excludes the left wall of the leftmost reservoir and the right wall of the rightmost reservoir, which are maintained at fixed temperatures $T_0$ and $T_e$. Accordingly, the species concentrations at these walls are given by $c_k\left(T_0\right)$ and $c_k\left(T_e\right)$, respectively, where $k=\{n,+,-\}$. The electric charge at the channel wall is assigned in terms of a constant surface charge density $q$. For a given $\zeta$ potential, the corresponding charge density is approximated by $q = \sqrt{8\epsilon k_B c_r T_r}\,\sinh\left(-e\nu\zeta/k_B T_r\right)$, as given by the non-linear Gouy-Chapman model for non-overlapping EDLs. A comparison between the numerically computed $\zeta$ potential (where the surface charge density was prescribed via the Gouy-Chapman model) and the input $\zeta$ potential values in the Gouy-Chapman model indicates that this provides a reasonable approximation for the range of values considered for Fig. 3 of the main text. The comparison, performed at a temperature of $(T_0 + T_e)/2$, is reported in Table S1 and demonstrates that the deviations always stay below 10% in the entire range $\kappa_0 h \in [0.1, 10]$. The most significant deviations between the prescribed and the numerically computed $\zeta$ potential values occur in the regime of strong EDL overlap. We refrained from employing a more complex analytical relationship between $\zeta$ and $q$ (such as presented in the Supplementary Material of [12]), since for strong EDL overlap, strictly, neither a fixed $\zeta$ potential nor a fixed surface charge density can be assumed. Rather than that, the boundary condition for the Poisson equation at the channel walls needs to be obtained from a charge regulation model that takes the specific surface chemistry into account [12].

The simulation software package Comsol Multiphysics 6.2 is used to carry out the numerical computations. Because of the mirror symmetry along the $z = 0$ plane, we consider only the top half of the domain for the computations. The channel length $l$ is set to $100h$, while the reservoir dimensions are taken to be $20h \times 10h$ (length$\times$height). We use triangular mesh elements with quadratic Lagrangian shape functions, employing mesh refinements close to the corners and the wall boundaries. The solutions are virtually independent of further mesh refinement at $\sim 8\times10^5$ mesh elements.

To compute $\mathcal{S}$ for an applied temperature difference of $\Delta T$, first, we calculate the electric current $I$ when the end walls of both reservoirs are maintained at $\phi = 0$. Thereafter, $\phi$ is varied for the rightmost boundary of the right-side reservoir (keeping $\phi = 0$ at the leftmost boundary of the left-side reservoir) until $I$ vanishes. At given $\kappa_0 h$ and $\zeta$ values, the particular value of $\Delta\phi$ for which $I = 0$ is achieved yields the Seebeck coefficient $\mathcal{S} = \Delta\phi / \Delta T$.

| $\kappa_0 h$ | $\zeta_{\mathrm{n}}$ ($\zeta = -100\,\mathrm{mV}$) [mV] | $\zeta_{\mathrm{n}}$ ($\zeta = -25\,\mathrm{mV}$) [mV] | $\zeta_{\mathrm{n}}$ ($\zeta = -10\,\mathrm{mV}$) [mV] |
|---|---|---|---|
| 10 | -99.54 | -24.88 | -9.91 |
| 0.1 | -109.96 | -27.25 | -10.70 |

**TABLE S1:** Numerically computed zeta potential $\zeta_{\mathrm{n}}$ for different $\zeta$ values as input parameters in the Gouy-Chapman model.

## §7: Effect of the applied temperature difference on the Seebeck coefficient

Here, we discuss the role of the applied temperature difference $\Delta T$ across the reservoirs on the induced concentration gradient of the charge carriers and, accordingly on $\mathcal{S}$. For this purpose, we consider two weak electrolyte systems: picric acid in 2-methoxyethanol and picric acid in 1, 2-ethanediol. The equilibrium dissociation constants of these electrolyte solutions display a nonlinear variation with temperature [S3,S4].

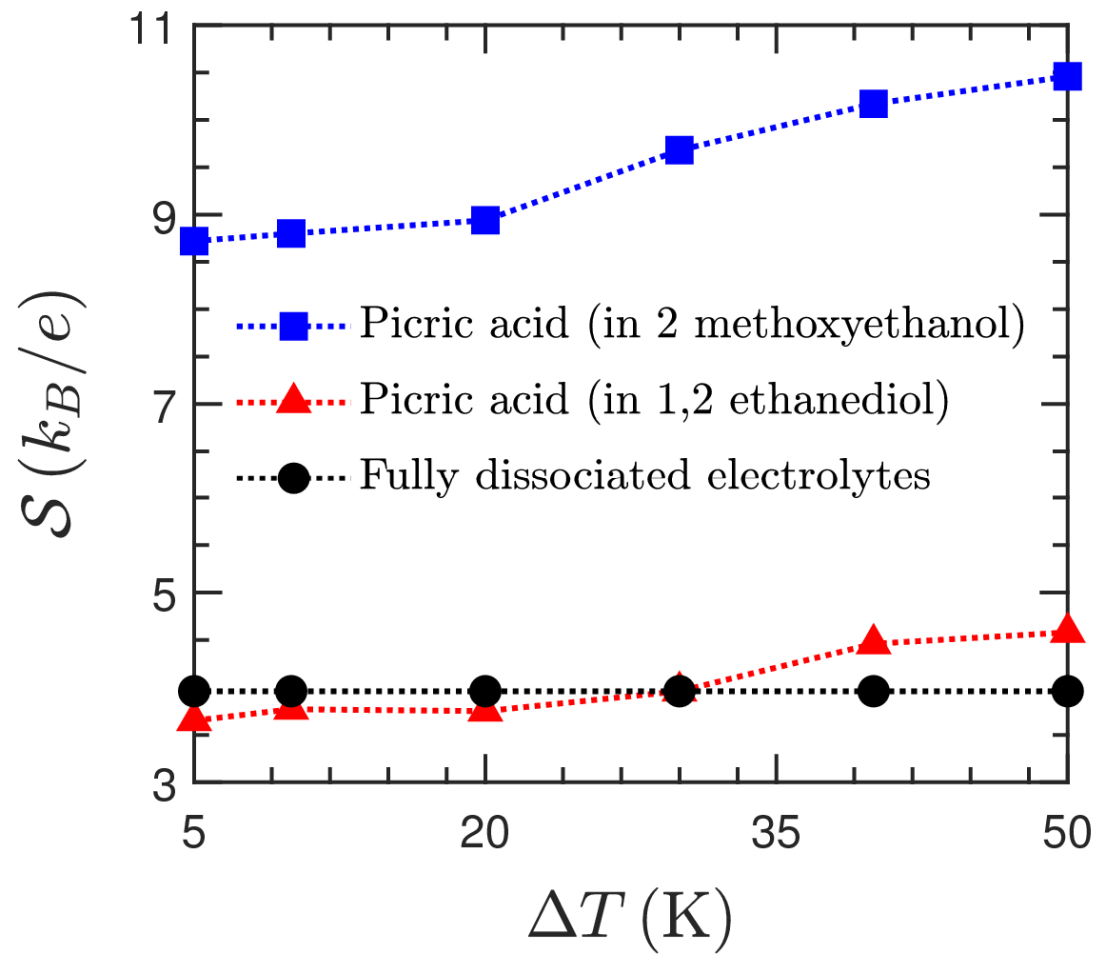

**FIGURE S2:** Dependence of the Seebeck coefficient (in units of $k_B/e$) for weak (blue and red symbols) and fully dissociated electrolytes (black circles) on the temperature difference $\Delta T$ across the reservoirs, where $\zeta = -100\,\mathrm{mV}$, $T_0 = 293\mathrm{K}$, and $\kappa_0 h = 0.1$.

In Fig. S2, the numerically computed Seebeck coefficient is plotted versus $\Delta T$, and a comparison with fully dissociated electrolytes is made. Because of the monotonic reduction of $\mathcal{K}_{eq}$ with temperature for picric acid in 2-methoxyethanol, the charge carrier concentration reduces with temperature. The higher $\Delta T$, the larger the axial concentration gradient, leading to an increase of the ion flux. Furthermore, the ion concentration originating from the temperature dependence of $D/\mu$ also reduces with increasing temperature. Both these effects augment the net ion flux, yielding a Seebeck coefficient that increases with $\Delta T$. On the other hand, for picric acid in 1,2-ethanediol, $\mathcal{K}_{eq}$ displays a non-monotonic variation with $T$ [S4], with a maximum around $303\,\mathrm{K}$. Therefore, with $T_0 = 293\,\mathrm{K}$, the charge carrier concentration increases with temperature up to $T = 303\,\mathrm{K}$ and reduces beyond $T > 303\,\mathrm{K}$. This causes the ion fluxes originating from the temperature dependencies of $\mathcal{K}_{eq}$ and $D/\mu$ to counteract for small $\Delta T$, resulting in a Seebeck coefficient lower than the one for fully dissociated electrolytes. Conversely, at higher $\Delta T$, the different fluxes point in the same direction and yield a thermovoltage higher than for fully dissociated electrolytes. From the above, it becomes clear that for confined weak electrolytes, $\mathcal{S}$ demonstrates a distinct dependence on the applied temperature difference between the reservoirs. This is in stark contrast to the characteristics of a fully dissociated electrolyte solution, where $\mathcal{S}$ remains independent of $\Delta T$, as demonstrated in Fig. S2.

[S1] P. Delahay and W. Vielstich, Kinetics of the Dissociation of Weak Acids and Bases—Application of Polarography and Voltammetry at Constant Current, J. Am. Chem. Soc. **77**, 4955 (1955).

[S2] R. R. Schroeder and I. Shain, Application of the potentiostatic method. Determination of the rate constant for the dissociation of acetic acid, J. Phys. Chem. **73**, 197 (1969).

[S3] G. Franchini, E. Ori, C. Preti, L. Tassi, and G. Tosi, A conductometric study of dissociation of picric acid in 2-methoxyethanol and 1,2-ethanediol from −10 to 80 °C, Can. J. Chem. **65**, 722 (1987).

[S4] G. C. Franchini, L. Tassi, and G. Tosi, The ethane-1,2-diol–2-methoxyethanol solvent system, J. Chem. Soc., Faraday Trans. 1 **83**, 3129 (1987).